\documentclass[12pt]{iopart}

\usepackage[T1]{fontenc}
\usepackage[dvips]{graphicx}
\usepackage{iopams}
\usepackage[applemac]{inputenc}
\usepackage{array}
\usepackage{booktabs}
\usepackage{slashbox}
\usepackage{setstack}

\newcommand{\ket}[1]{\vert #1 \rangle}

\begin{document}
\title{Production of Sodium Bose--Einstein condensates in an optical dimple trap}
\author{D Jacob, E Mimoun, L De Sarlo, M Weitz\dag, J Dalibard and F Gerbier}
\address{Laboratoire Kastler Brossel, ENS, UPMC-Paris 6, CNRS; 24 rue Lhomond, 75005 Paris, France}
\address{\dag Institut für Angewandte Physik, Universität Bonn, Wegelerstr. 8, 53115 Bonn, Germany}
\date{\today}
\submitto{\NJP}

\begin{abstract}

We report on the realization of a sodium Bose--Einstein condensate (BEC) in a combined red-detuned optical dipole trap, formed by two beams crossing in a horizontal plane and a third, tightly focused dimple trap propagating vertically. We produce a BEC in three main steps: loading of the crossed dipole trap from laser-cooled atoms, an intermediate evaporative cooling stage which results in efficient loading of the auxiliary dimple trap, and a final evaporative cooling stage in the dimple trap. Our protocol is implemented in a compact setup and allows us to reach quantum degeneracy even with relatively modest initial atom numbers and available laser power.
\end{abstract}

\section{Introduction}

The preparation of degenerate atomic quantum gases is interesting from both a fundamental and an applied point of view. On the one hand, the unprecedented level of control on these systems allows one to study quantum many-body phenomena in the absence
of perturbing effects unavoidable in solid-state systems \cite{BlochZwergerDalibard}. On the other hand, degenerate gases are a promising starting point to reliably produce highly entangled states, which could pave the way for a new generation of atom-based quantum sensors (see \cite{Bouyer_HeisenbergLimitedSpectroscopy,Gross_NonLinearInterferometry} and references  therein).

In view of the sensitivity of these strongly correlated states to the perturbations caused by magnetic fields fluctuations, experimental schemes in which evaporative cooling is performed without the use of external magnetic fields are particularly interesting. These so-called ``all-optical evaporation'' schemes rely on far off-resonant optical dipole traps. They have been developed by several groups to produce Bose--Einstein condensates (BEC) of various atomic species, in particular alkali atoms (Rb \cite{Barrett_AllOpticalBECRb,Kinoshita_AllOpticalBECcompressibleCDT,Weitz_CO2trapBEC}, Li \cite{Granade_AllOpticalFermiGas}, Cs \cite{Grimm_CsBEC} and Na \cite{Lett_ProductionAllOpticalNaBEC}). In such all-optical setups, the  trapping potential is almost independent of the internal state, opening the route to the study of spinor condensates \cite{Ketterle_SpinorCondensates}. 

Experiments relying on all-optical setups are based on a common experimental scheme: Laser-cooled atoms are first loaded into an optical dipole trap and then evaporatively cooled by lowering the trapping laser intensity and thus the trap depth. In this paper we discuss how to optimize these two steps for producing an all-optical BEC of sodium atoms, starting with relatively modest atom numbers and laser powers.

The first issue, dealing with the transfer from the magneto-optical trap (MOT) to the optical dipole trap, has extensively been studied (see e.g. \cite{Kuppens_LoadingAnODT}). Laser cooling forces and light-assisted losses can be strongly modified by the presence of the dipole trap potential. The size and the depth of the trapping potential have to be adapted to the size, density and temperature of the MOT. A convenient configuration is a laser trap consisting of two crossed gaussian beams \cite{Chu_CrosseDipoleTrapEvapCooling}, as in our experiment. The loading of the trap then occurs in two steps: Atoms are first captured in both arms of the trap, and then start filling the crossing region by ``free evaporation'' once the near-resonant cooling beams have been turned-off \cite{Barrett_AllOpticalBECRb}. We will describe our procedure to optimize the atom number $N$ in the crossing region and the temperature $T$ at the end of the free evaporation to get a high phase-space density $\mathcal{D}$ and a large collision rate $\gamma_{\rm coll}$. We recall that $\mathcal{D}=N\left(\hbar\omega/k_{\rm B}T\right)^3$ and $\gamma_{\rm coll}\propto N/\omega^3T$ for a Boltzmann gas in a harmonic trap, with $\hbar$ the Planck constant, $k_{\rm B}$ the Boltzmann constant, and $\omega$ the average trapping frequency. For a given beam size, we find that the optimal trap depth is different for loading and free evaporation, and propose that trap-induced light-shifts on the cooling transition are the physical mechanism behind this observation.

The second issue is related to the efficiency of evaporative cooling. In these respect, optical traps differ in several aspects from magnetic traps. In magnetic traps, evaporative cooling takes place in the so-called runaway regime, where the elastic collision rate  $\gamma_{\rm coll}$ and evaporation efficiency stay constant or even increase with time \cite{KetterleDruten_Evaporativecoolingoftrappedatoms}. In optical traps, this regime is not easily reachable because the trap depth and trap confinement both increase with the trapping laser power. In practice, decreasing the trap depth to force evaporation results in a looser confinement, so that the collision rate can decrease even if the phase-space density increases. Solutions involving modification of the trapping potential have been demonstrated to resolve this issue. For example a dynamical change of the beam size using a zoom lens allows one to maintain constant confinement while reducing the trap depth \cite{Kinoshita_AllOpticalBECcompressibleCDT}, thus preserving a high collision rate during evaporation. Runaway evaporation in an optical trap can also be obtained, by using an additional expelling potential independent of the trapping laser (gravity or ``pulling'' laser) in order to decouple trap confinement and potential depth \cite{Chin_AcceleratingEvaporativeCooling,Clement_AllOpticalRunAwayEvaporation}. A third solution, based on the addition of a tighter ``dimple'' potential \cite{Grimm_CsBEC,Porto_CombineMagneticOpticalPotential}, has been realized and characterized theoretically \cite{Foot_Cesium,Foot_Dimple,Comparat_OptimLargeBECs}. This solution, which is the one investigated in this paper, leads to a two-step evaporation sequence: After the loading of a larger trap, atoms are first transferred by cooling into the ``dimple'' trap, then further cooled down in this trap. The major advantages of this technique are its relative technical simplicity (as compared, for instance, to a ``zoom-lens'' method), the increase of phase-space density during the transfer, and the high efficiency of the second evaporation step due to the high confinement in the ``dimple'' trap. Here we describe the first application of this technique to $^{23}$Na. Starting with $3\times10^5$ trapped atoms, a Bose--Einstein condensate of $\sim10^4$ $^{23}$Na atoms is produced after $\sim2~$s evaporation time.

The paper is organized as follows. In section \ref{ExperimentalSetup}, we give an overview of our experimental setup. In section \ref{Loading}, we investigate the loading of a dipole trap from a MOT of sodium atoms and study how the compression of the trap after the atom capture improves the initial conditions for evaporative cooling. We then present in section \ref{evaporation} how evaporative cooling works in presence of the auxiliary dimple trap, detailing its filling dynamics, and the last evaporative cooling stage to reach Bose--Einstein condensation.

\section{Experimental Setup}
\label{ExperimentalSetup}

\subsection{Laser cooling}
\label{LaserCooling}

Our experiment starts with a sodium magneto-optical trap (MOT) capturing approximately $10^7$ atoms in $10~$s from a vapor whose pressure is modulated using light-induced atomic desorption \cite{mimoun_LIAD}. After the MOT is formed, a far off-resonant dipole trap is switched on (see subsection \ref{TrapLaserConfiguration}). The detunings and powers of both the cooling (tuned to $3S_{1/2},F=2\rightarrow3P_{3/2},F'=3$ transition) and repumping (tuned to $3S_{1/2},F=1\rightarrow3P_{3/2},F'=2$ transition) lasers are modified in order to optimize the trap loading. During a first ``dark MOT'' phase \cite{Ketterle:1993fk}, we lower the power of the repumping laser in about $100~$ms, from $I_{\rm rep}=300~\mu$W.cm$^{-2}$ to $I_{\rm rep}=10~\mu$W.cm$^{-2}$ per beam while keeping the magnetic gradient on. This reduces the loss rate due to light-induced collisions by limiting the population of the excited states \cite{Kuppens_LoadingAnODT}. We keep the cooling laser intensity at the same value as for MOT loading, $I_{\rm cool}=0.9~$mW.cm$^{-2}$ per beam, which corresponds to one sixth of the saturation intensity ($I_{\rm sat}=6.3~$mW.cm$^{-2}$). During this ``dark MOT'' phase, both the spatial density in the dipole trap and the temperature increase. We then apply a $30~$ms-long ``cold MOT'' phase, where the cooling beam detuning is shifted from $\delta_{\rm cool}\approx-\Gamma$ to $\delta_{\rm cool}\approx -3.8\Gamma$ ($\Gamma/2\pi \approx 10~$MHz is the natural linewidth). The temperature of the atoms after this cooling sequence is around $50~\mu$K.

\subsection{Trapping lasers configuration}
\label{TrapLaserConfiguration}

The far off-resonant dipole trap results from the combination of three beams, two forming a crossed dipole trap (CDT) in the horizontal $x-y$ plane and a tightly focused one propagating vertically along the $z$ axis (see figure \ref{Scheme2}(a)), which we refer to as ``dimple trap'' (dT). The CDT is derived from a $40~$W fiber laser (IPG Photonics) at $1070~$nm. This trap is formed by folding the beam onto itself at an angle $\theta\simeq45^{\circ}$ in the horizontal plane. At the crossing point, both arms have a waist $w_{\rm CDT}\approx42~\mu$m. We control the laser power using a motorized rotating waveplate (OWIS GmbH) followed by a Glan-Taylor polarizer (bandwidth $\sim10~$Hz), and a control input on the current in the laser pump diodes (bandwidth $\sim50~$kHz). The waveplate is used for coarse reduction of laser power by changing the amount of light transmitted by the polarizer, whereas the current control is used at the end of the evaporation ramp (low laser powers) and for fast servo-control of the intensity to reduce fluctuations. Combining both servo loops, we can control the laser power from its maximal value ($P_{\rm CDT}\approx 36~$W) down to  $\approx 100~$mW. We can switch off the trapping potential to an extinction level greater than $90~\%$ in less than $10~\mu$s using the laser current input. We use motorized mirrors (Agilis, Newport Corporation) for alignment. Special care is taken to ensure the orthogonality of the polarization of both arms, realized by the insertion of a $\lambda/2$ waveplate that is positioned with a precision $\lesssim0.5^{\circ}$. A misalignement of only $1^{\circ}$ results in a measurable heating of the sample \cite{mimoun_LIAD}.

The auxiliary dimple trap is produced using a $500~$mW laser (Mephisto-S, InnoLight GmbH) at $1064~$nm. As sketched on figure \ref{Scheme2}(a), the beam propagates vertically, and crosses the CDT with a waist $w_{\rm dT}\approx8~\mu$m. The laser beam is transmitted through a single-mode optical fiber, and focused to a waist size $w_{\rm dT}$ using a custom-made microscope objective (CVI Melles Griot, ${\rm NA}\gtrsim0.3$). An acousto-optic modulator placed before the fiber allows us to control the intensity and to quickly switch off the dT-beam. 

To fix the notations that will be used in the following, we give here the expressions of the dipole trap potentials. The expression of the CDT potential is given by
\begin{equation}
V_{\rm CDT}(x,y,z)=-\frac{V^0_{\rm CDT}}{2} \left[ \frac{\rme^{-2(x^2+z^2)/w(y)^2}}{\left(w(y)/w_{\rm CDT}\right)^2}+\frac{\rme^{-2(u^2+z^2)/w(v)^2}}{\left(w(v)/w_{\rm CDT}\right)^2}\right], \\
\label{CDTpotential}
\end{equation}
with $w(y)=w_{\rm CDT}\sqrt{1+y^2/y_R^2}$ and with $y_R$ the Rayleigh length $y_R=\pi w_{\rm CDT}^2/\lambda\approx5.2~$mm. We have also introduced the rotated coordinates: $(u,v) = (x\cos(\theta)+y\sin(\theta),-x\sin(\theta)+y\cos(\theta))$.
The expression of the dT potential is given by
\begin{equation}
V_{\rm dT}(x,y,z)=-V^{0}_{\rm dT}\,\rme^{-2(x^2+y^2)/w_{\rm dT}^2},
\end{equation}
neglecting the confinement of the dT along the $z$-axis, always negligible compared to the vertical confinement of the CDT. Typical trapping frequencies and potential depths are given in table \ref{tableFreq}.

\begin{table}
\centering
\begin{tabular}{|c| >$c<$ | >$c<$ | >$c<$ | >$c<$|}
\toprule
Dipole Trap &\omega^{x}/2\pi ~({\rm kHz}) & \omega^{y}/2\pi~({\rm kHz})  & \omega^{z}/2\pi~({\rm kHz})  & V_0/k_{\rm B} ~({\rm mK}) \\
\midrule
CDT & 2.5 & 4.5 & 5.1 & 1.2\\
dT & 3.7 & 3.7 & 0.021 & 0.10\\
\bottomrule
\end{tabular}
\caption{\label{tableFreq} Trapping frequencies and trap depths at $P_{\rm CDT}\approx 36~$W and $P_{\rm dT}\approx 200~$mW for the crossed dipole trap (CDT) and the dimple trap (dT), respectively.}
\end{table}

In figure \ref{Scheme2}(c), we give schematically the temporal evolution of the powers of the two lasers during the experimental sequence. This time evolution is optimized for loading and evaporation, as explained in sections \ref{Loading} and \ref{evaporation}.

\subsection{Imaging}

\begin{figure}[H]
\includegraphics{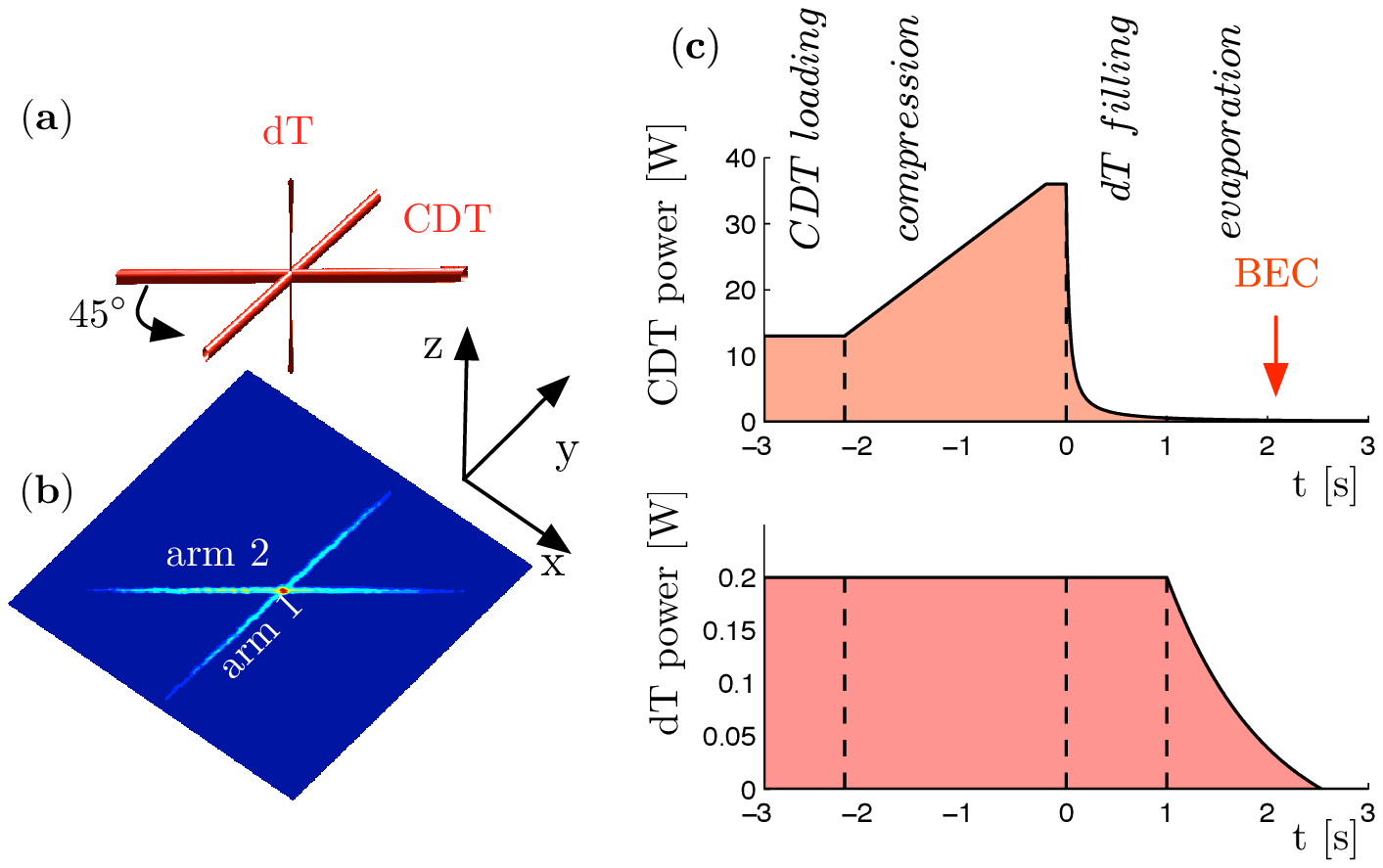}
\caption{(a) Sketch of the laser geometry showing the crossed dipole trap (CDT) propagating in the horizontal plane and the dimple trap (dT) propagating vertically. (b) Fluorescence image of the atoms trapped in the CDT taken after short  time-of-flight . The thermal equilibrium state in such a potential has a characteristic spatial structure: two elongated ``arms'' and a denser crossing region. (c) Evolution of the powers of the crossed dipole trap and the dimple trap during the sequence. The first step corresponds to the loading of the CDT from a ``cold-MOT'' phase, followed by a compression that helps to fill the central trapping region. The next step consists in evaporatively cooling the CDT and results in the filling of the dimple trap. The last step is evaporative cooling in the dimple trap, which leads to Bose--Einstein condensation.}
\label{Scheme2}
\end{figure}

We monitor the time evolution of the trapped cloud using both fluorescence and absorption imaging \cite{Ketterle_HouchesLectureNotes}. Fluorescence light is captured by the same high numerical aperture microscope used to focus the dT. The photons are collected on a low-noise charge-coupled device camera (PIXIS, Princeton Instruments). In figure \ref{Scheme2}(b), we show a typical fluorescence image. We typically observe the atoms after a time-of-flight $t_{\rm ToF}=0.1~$ms during a short pulse  ($t_{\rm mol}=50~\mu$s) performed with the six cooling and repumping beams.

Absorption images are recorded with a vertically propagating resonant probe beam. It is better suited for the analysis of the central denser part of the trapped cloud but not very precise for the arms of the CDT. Indeed, the regions corresponding to the CDT arms display low optical densities ($<0.1$) only slightly above the noise level ($\sim0.04$, limited by residual fringes on the background). Atom counting in the arms of the CDT is thus more accurate using fluorescence images. 

\section{Loading and free evaporation in the Crossed Dipole Trap}
\label{Loading}

\subsection{Dipole trap loading dynamics}
\label{Diagnostic}

\begin{figure}
\includegraphics[scale=1]{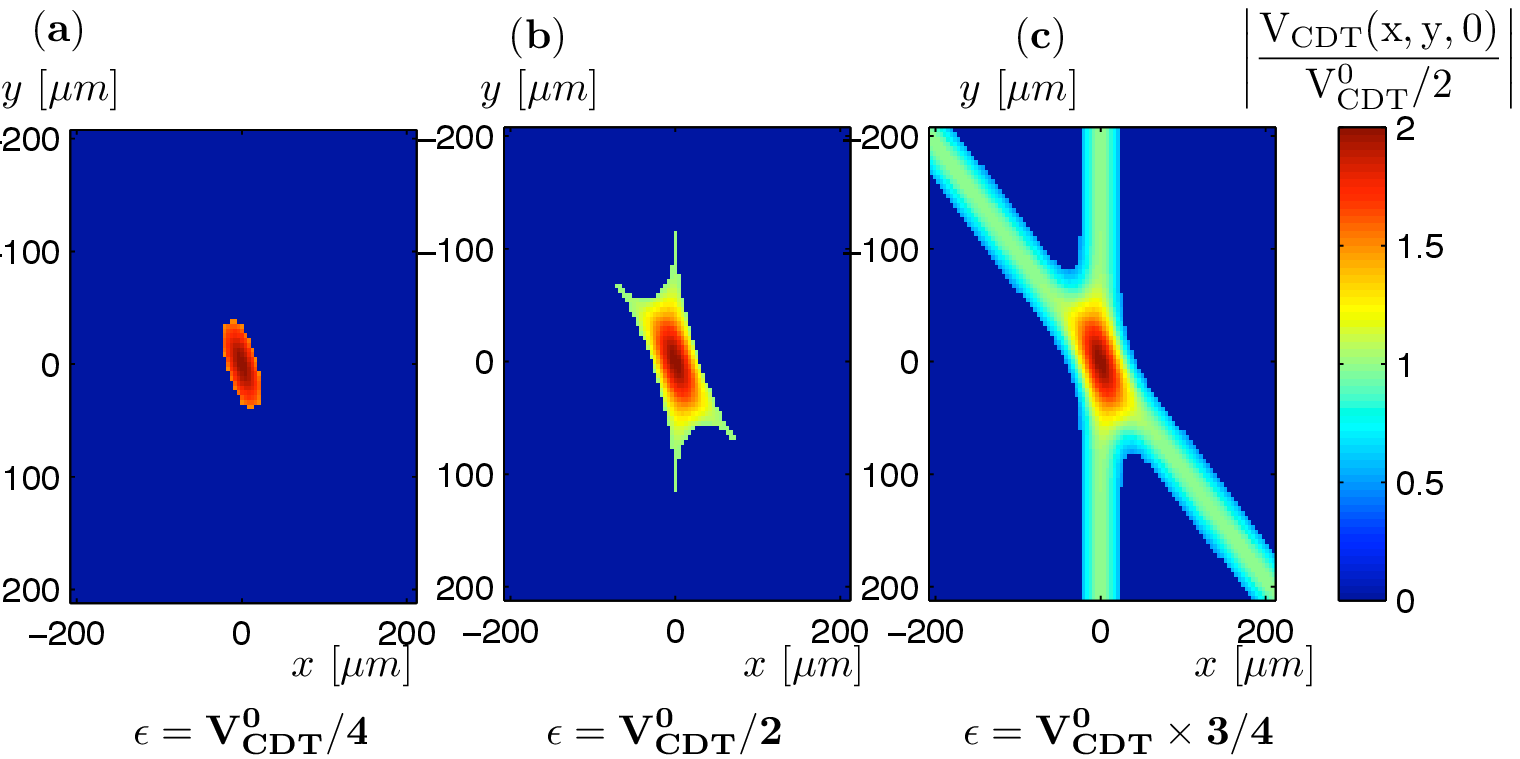}
\caption{CDT potential in the $z=0$ plane, truncated at an energy $\epsilon=V^0_{\rm CDT}/4$ (a), $\epsilon=V^0_{\rm CDT}/2$ (b), and $\epsilon=V^0_{\rm CDT}\times3/4$ (c).}
\label{IntegrandCDT}
\end{figure}

We can distinguish two stages in the dynamics of the trap loading. At first, during the MOT/CDT overlap period, atoms are captured mainly in the arms of the CDT without a notable enhancement of the density in the crossing region. In the second phase that follows the extinction of the MOT beams, which we call ``free evaporation'', the hottest atoms leave the arms and the remaining ones fill the crossing region through thermalization. The quantity of interest is the number of atoms in the central region $N_{\rm C}$, which corresponds approximately to the number of atoms with an energy comprised between $-V^0_{\rm CDT}$ and $-V^0_{\rm CDT}/2$ (as defined in equation (\ref{CDTpotential})). We show in figure \ref{IntegrandCDT} the potential $V_{\rm CDT}$ in the $z=0$ plane, truncated at three different energy levels. One can see that atoms having energies lower than $V^{0}_{\rm CDT}/2$ explore only the central region, as expected. This dense part is the relevant component that matters for further evaporative cooling.
Although both trapping lasers (CDT and dT) are turned on simultaneously, the CDT is much deeper than the dimple trap, the latter playing a negligible role during this initial stage. In this section, we discuss auxiliary experiments where the dT is absent.

In order to understand the loading dynamics during the first stage, we give a brief overview on the relevant mechanisms (see \cite{Kuppens_LoadingAnODT} for a detailed analysis). The loading rate of atoms in the CDT is proportional to the probability of an atom to be trapped by the dipole potential and to the atomic flux in the CDT/MOT overlap region. The first term corresponds to the damping of the velocity of an atom when it crosses one arm of the CDT, leading to a reduction of its total energy below the CDT potential depth. The second term is proportional to the spatial density and the average velocity of the atoms in the MOT, and thus depends on the temperature of the atoms. The relevant parameters for optimizing the loading rate, namely the atomic density and the temperature, can be adjusted by the ``dark MOT'' and ``cold MOT'' phases (see subsection \ref{LaserCooling}). The presence of the dipole potential changes locally the cooling properties, due to the light shifts induced by the CDT laser beams. During this phase, atom accumulation in the trap crossing region is limited by light-assisted inelastic collisions, such as radiative escape.

In the second stage, after the MOT light is extinguished, the trapped atoms thermalize and the sample cools down by evaporative cooling (at a fixed potential depth). Atoms concentrate in the crossing region and the phase-space density increases substantially as compared to the MOT \cite{Barrett_AllOpticalBECRb}.

We have experimentally tested CDT configurations with different beam sizes $w_{\rm CDT}$ (from $30~\mu$m to $50~\mu$m). A larger beam size helps to trap more atoms during the capture stage due to higher overlap volume. However, at a given available power, larger beams imply a weakening of the trap stiffness, which in return penalizes the thermalization after capture. The data presented in this paper are taken with a beam waist $w_{\rm CDT}\approx42~\mu$m. We obtain very similar results for $w_{\rm CDT}\approx35~\mu$m, but with different optimal powers at each stage. In the next subsection, we will concentrate on the optimization of the laser power to find the optimal trap depth for filling the center region. 

\subsection{Optimization of CDT loading}
\label{LoadingExperiment}

\begin{figure}
\includegraphics[scale=1]{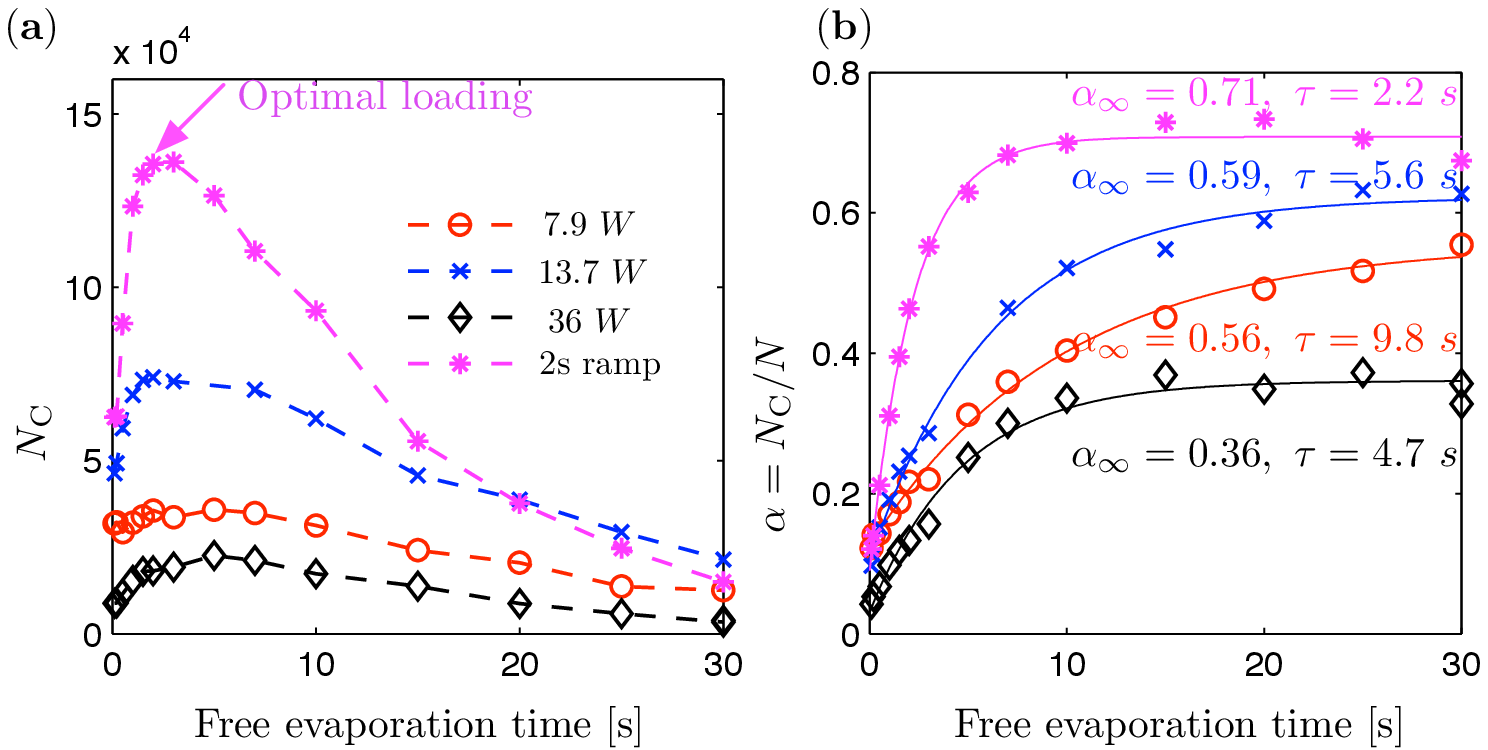}
\caption{Evolution of the atom number in the center of the CDT (a) and the loading ratio $\alpha=N_{\rm C}/N$ (b) in four different loading situations: low power (circles), highest power (diamonds), $\sim1/3$ of maximum power (crosses), ramping up in $2~$s after loading at low power (stars). The loading ratio $\alpha(t)$ is fitted with the function $a(1-\rme^{-t/\tau})+b$. The results of the fit $\tau$ and $\alpha_{\infty}=a+b$ are indicated in (b).}
\label{ComparaisonRampNormal_Fluo}
\end{figure}

In order to characterize the filling dynamics of the crossing region, we define the filling factor $\alpha=N_{\rm C}/N$ as the fraction of atoms in this region relatively to the total number of atoms in the dipole trap. Images as that in figure \ref{Scheme2}(b) are processed with a multi-component fitting routine that extracts the temperature, the density, the total atom number $N$ and $N_{\rm C}$. Details about the fitting procedure are presented in \ref{DataAnalysis}. The results of the optimization of the CDT power are presented in figure \ref{ComparaisonRampNormal_Fluo}, where we plot the evolution of atom number $N_{\rm C}$ and filling factor $\alpha$ with time. We fit the function $\alpha(t)=a(1-\rme^{-t/\tau})+b$ to our data.

\begin{figure}
\includegraphics[scale=1]{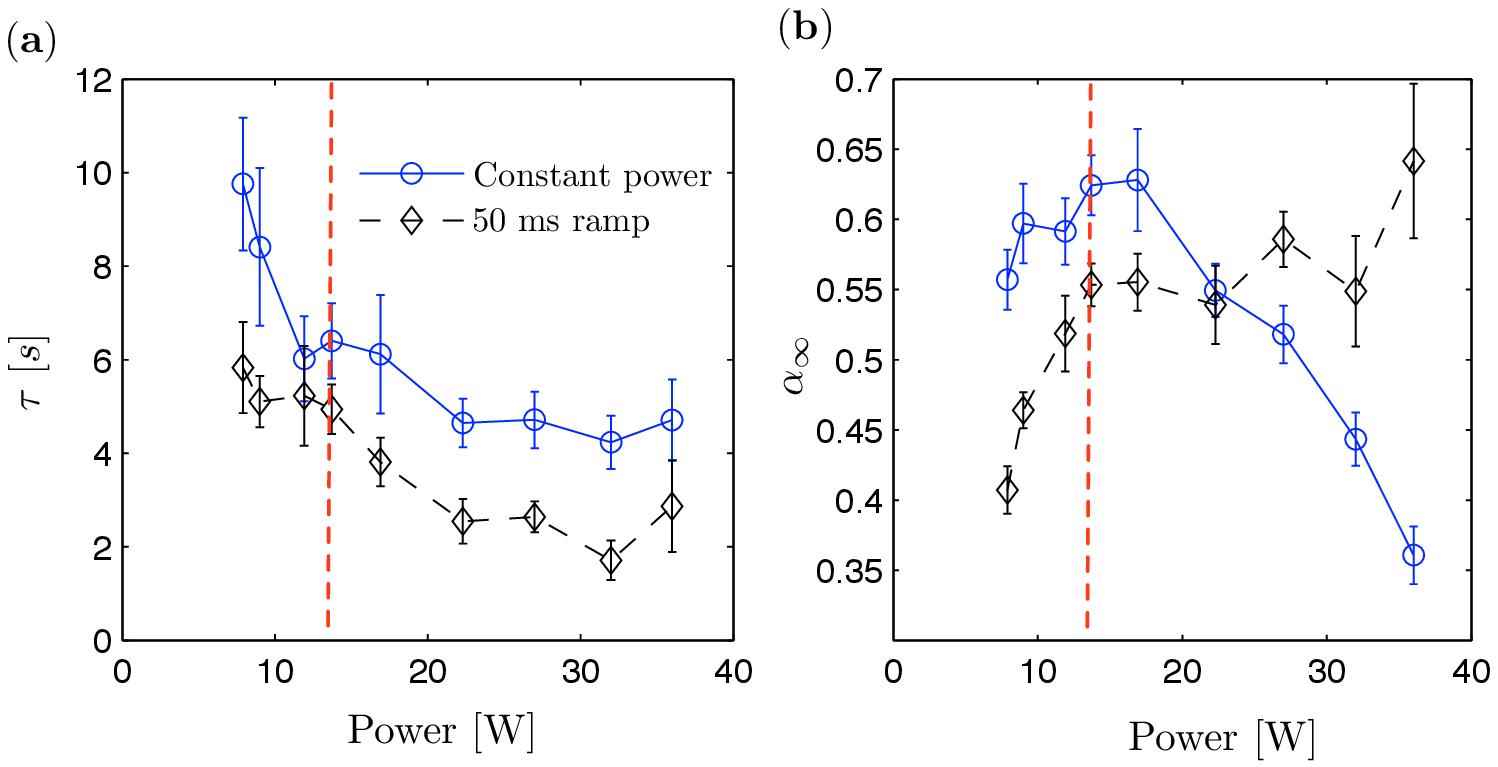}
\caption{(a) Filling time $\tau$ and (b) center filling fraction $\alpha_{\infty}$ for CDT in two different situations: the solid curve shows the results of the experiments where the CDT laser is hold at any time at the same power. The dashed curve denotes the compression experiments where the power starts at $P_{\rm CDT}=13.7~$W in the ``cold MOT'' phase and is ramped up in $50~$ms to the final value indicated after switching off the molasses beams. The error-bars correspond to $90\%$ confidence bounds on the fit coefficients $\tau$ and $\alpha_{\infty}$. For $P_{\rm CDT}=13.7~$W, both curves should intersect as the experimental sequence is the same. The observed difference indicates systematic variations between different experimental runs, probably due to dipole trap pointing fluctuations and total atom number variations. The vertical dashed line corresponds to the optimal power $P_{\rm CDT}=13.7~$W for the ``cold MOT'' phase.}
\label{AnalyzeLoading_Fluo}
\end{figure}

We look first at a situation in which ``free evaporation'' occurs at constant CDT power, keeping the same power during the free evaporation phase as during the capture stage. We report in figure \ref{ComparaisonRampNormal_Fluo} the evolution of $N_{\rm C}$ and $\alpha$ with time for three different powers ($P_{\rm CDT}=7.9, 13.7$ and $36~$W). The values of the loading time $\tau$ and the asymptotic value $\alpha_{\infty}=a+b$ of the filling factor obtained from the fit are shown in figure \ref{AnalyzeLoading_Fluo}. We find an optimal power $P_{\rm CDT}=13.7~$W that maximizes both the number of atoms $N_{\rm C}$ and the stationary filling fraction $\alpha_{\infty}$.

In a second set of experiments, the CDT is kept at constant $P_{\rm CDT}=13.7~$W during the ``cold MOT'' phase, and ramped up in $50~$ms to another value just after switching off the resonant lasers. As shown in figure \ref{AnalyzeLoading_Fluo}, ramping up the power to the maximum available power results in quicker loading of the central region ($\simeq2~$sec) and better filling ratio ($\alpha_{\infty}\simeq0.6$), the best values being apparently limited by the available laser power. A slower, linear power ramp to $P_{\rm CDT}=36~$W in $2~$sec (also shown in figure \ref{ComparaisonRampNormal_Fluo}), leads to a slightly better loading ratio ($\alpha_{\infty}\simeq0.7$), and also a slightly lower temperature, which altogether results in a higher number of atoms (about twice as many atoms in the center of the CDT as compared to the loading at constant $P_{\rm CDT}=13.7~$W). This particular ramp provides the best starting point we could achieve for the evaporative cooling stage.
 
The results of the two series of experiments show the existence of an optimal power $P^{\rm opt}_{\rm CDT}=13.7~$W for the loading of the atoms during the period in which the MOT and the CDT are simultaneously present. We interpret this observation in the following way. The CDT laser exerts different light shifts on the various hyperfine states in the ground ($3s$) and excited ($3p$) manifolds. These differential light shifts can perturb the laser cooling dynamics in the CDT region and thus degrade the capture efficiency. For instance, if we take the $\ket{g}=\ket{F=2,m_F=2}~\rightarrow~\ket{e_3}=\ket{F'=3,m_F=3}$ transition and a $\pi$-polarized CDT laser, we obtain that near the trap bottom, the laser detuning changes according to
\begin{equation}
\delta_{33}=\omega_{L}-\omega_{33}+\alpha_{33}I,
\end{equation}
with $\omega_{L}$ the cooling laser frequency, $\omega_{33}$ the ``bare'' transition frequency, $I=2P_{\rm CDT}/\pi w_{\rm CDT}^2$ the intensity of the CDT laser \footnote{For the calculation we use the data from the NIST atomic spectra database \cite{NIST_spectra} and consider the $3s~\rightarrow~3p$, $3p~\rightarrow~3d,4d$ transitions (see also \cite{Viering_thesis}).}. For sodium, we find $ \alpha_{33}/2\pi\approx 27~$Hz.cm$^2$/W. For our optimal cooling beam detuning $\delta_{\rm cool}=\omega_{L}-\omega_{33}\approx -3.8\Gamma$ (see section \ref{LaserCooling})  we obtain that the detuning on the cooling transition vanishes when $I \approx \vert \delta_{\rm cool}\vert/\alpha_{33}\approx 1.4\times10^6~{\rm W.cm^{-2}}$. Experimentally, the optimum $P^{\rm opt}_{\rm CDT}=13.7~$W corresponds to $I^{\rm opt}=4.7\times10^5~{\rm W.cm^{-2}}$, close to the value calculated above, and a change of detuning from $-3.8\Gamma$ to $\delta_{33}\approx-2.5\Gamma$. We reached a very similar optimum in another set of experiments with $w'_{\rm CDT}=35~\mu$m, where we found an optimum power $P'^{\rm opt}_{\rm CDT}=10~$W corresponding to $I'^{\rm opt}=5.2\times10^5~{\rm W.cm^{-2}}$ and a comparable final detuning $\delta_{33}\approx-2.4\Gamma$.

One could think that tuning the cooling beam frequency further than $-3.8\Gamma$ on the red side of the $\ket{g}~\rightarrow~\ket{e_3}$ transition could help to mitigate the effect, thus increasing the optimal power and ultimately the number of atoms captured. However, two separate effects work against this strategy. First, this compensation is efficient only near the trap bottom and not across the whole trapping region. Second, it brings the MOT-beams closer to resonance with neighbouring transitions that can shift in opposite ways. For example the $\ket{g}=\ket{F=2,m_F=2}~\rightarrow~\ket{e_2}=\ket{F'=2,m_F=2}$ transition has an intensity dependance $\delta_{22}=\omega_{L}-\omega_{22}-\alpha_{22}I$, with $\omega_{22}$ the corresponding frequency and $\alpha_{22}/2\pi\approx 16~$Hz.cm$^2$/W. The latter effect is limiting for $^{23}$Na, which has an hyperfine structure splitting $\omega_{33}-\omega_{22}$ much smaller than heavier alkalis ($^{87}$Rb and $^{133}$Cs).

\section{Two-stage evaporation}
\label{evaporation}

\subsection{Evaporation in the crossed dipole trap alone}

As pointed out in the introduction, lowering the laser intensity to reduce the trap depth for evaporation is inevitably accompanied by a reduction of trap stiffness (near the trap bottom, the trap frequency $\omega$ is proportional to $\sqrt{P}/w$), unlike in magnetic traps where the depth and confinement are independent. The resulting decrease in density and collision rate can make the cooling due to evaporation stop at low laser power, and this is precisely what is observed in our experiment. For a harmonic trap the classical phase-space density is given by $\mathcal{D}=N\left(\hbar\overline{\omega}/k_{\rm B}T\right)^3$, where $\overline{\omega}$ stands for the mean trapping frequency. In a simple model where the evaporation parameter $\eta=V_0^{\rm CDT}/k_{\rm B}T$ is assumed constant and where losses are neglected, the gain in phase-space density when the trap depth is lowered from $V_0^{\rm CDT}$ to $V_0^{\rm CDT}/r$ ($r>1$ is the reduction factor) is given by \cite{OHara_ScalingLaws}
\begin{equation}
\label{model}
\mathcal{D}=\mathcal{D}_0 r^{\beta},~~~ \beta = {\frac{3}{2}\frac{\eta^2-7\eta+11}{\eta^2-6\eta+7}}.
\end{equation}

The starting point in our experiment (about $3\times10^{5}$ atoms at $T\simeq100~\mu$K) corresponds to $\eta\approx10$ and a phase-space density $\mathcal{D}_0\sim10^{-4}$. According to equation (\ref{model}), evaporating with a reduction factor $r=200$ leads to a final phase-space density of $\sim0.2$. In our experiment, the laser power is ramped down during evaporation according to
\begin{equation}
\label{CDTramp}
V_{\rm CDT}(t)=V^0_{\rm CDT}\left(1 + t/\tau_{\rm evap}\right)^{-\alpha_{\rm evap}},
\end{equation}
where the parameters $\tau_{\rm evap}=30~$ms and $\alpha_{\rm evap}=1.2$ were optimized empirically. Even after optimization we have not been able to achieve a final phase-space density greater than $\sim10^{-2}$ in the CDT alone \footnote[1]{The measured phase-space density is lower than the prediction of equation (\ref{model}). This should be attributed to the crudeness of the model underlying this equation. In particular, three-body losses, which are important at the densities present in the CDT, are not accounted for.}. We also observe that the collision rate after $\sim1~$s is lower than $10~$s$^{-1}$, and that evaporative cooling stops near this point. Such a collision rate is too low to maintain efficient thermalization and sustain the cooling process.

\subsection{Evaporation in the dimple trap}

The decrease of evaporation efficiency presented in the previous subsection is caused by the relation between trap depth and confinement strength inherent to optical traps. To circumvent this problem, one needs to break this relation. To this aim, we have added a tight ``dimple'' to the initial trap \cite{mimoun_LIAD,Grimm_CsBEC,StamperKurn_ReversibleBEC,Davis_GrowthDynamicBECdimple}. We turn on the auxiliary dT together with the main CDT, but keep it at constant power $P_{\rm dT}=200~$mW during the CDT ramp \footnote[3]{Keeping the dT power high causes no modification for the CDT loading and compression.}. With the ``dimple'' addition, the atoms feel a more and more confining potential as they cool, in stark contrast to the situation in the CDT alone. In our experiment, we take advantage of the dissipative nature of evaporative cooling to fill such a ``dimple'' trap (dT). As the atoms in the CDT are evaporatively cooled, they get progressively trapped in the stiffer potential which results in a substantial increase in spatial density \cite{Grimm_CsBEC, Foot_Dimple}. Since the temperature remains the same, this translates into a huge boost in the phase-space density. This is markedly different from an adiabatic trap compression which increases spatial density but leaves the phase-space density unchanged \cite{Ketterle_AtomCoolingByTimeDependentPot}. After $1~$s evaporation in the CDT, the dT power is reduced to provide the final stage of evaporative cooling (see figure \ref{Scheme2}(c)).

\begin{figure}[H]
\centering
\includegraphics[scale=1]{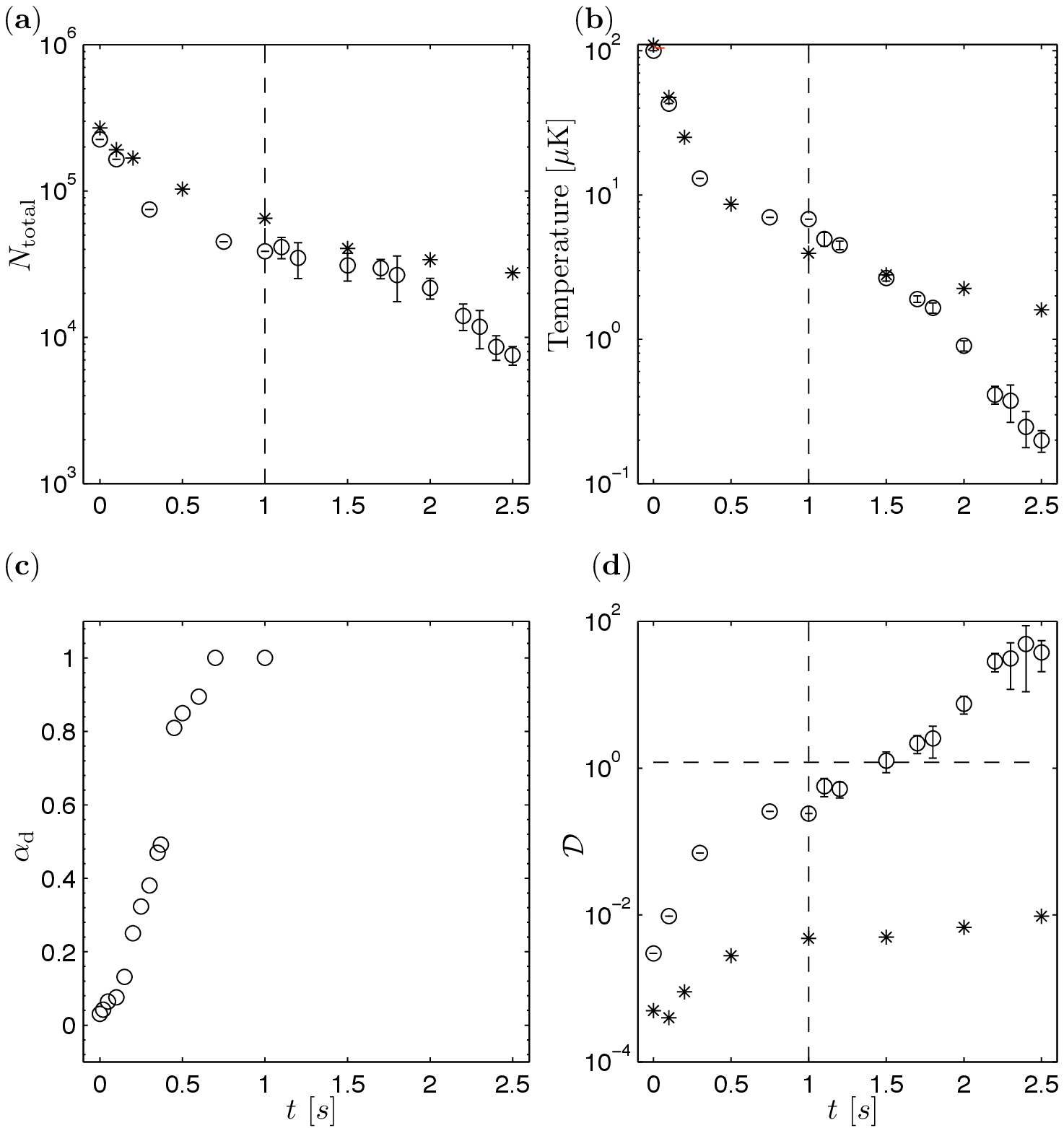}
\caption{Evaporative cooling trajectories in the combined trap (CDT and dT) (circles) and in the CDT alone (stars). We show the time evolution of atom number $N$ (a), temperature $T$ (b), dimple filling $\alpha_{\rm d}$ (c) and phase-space density $\mathcal{D}$ (d).}
\label{evapDimpleShorterNoSimul}
\end{figure}

The plots in figure \ref{evapDimpleShorterNoSimul} summarize the evaporation dynamics. We show atom number $N$, temperature $T$, dimple trap filling $\alpha_{\rm d}=N_{\rm d}/N$ where $N_{\rm d}$ is the number of atoms present in the dT, and phase-space density $\mathcal{D}$ during the ramp, and compare it to the evaporation without dimple trap\footnote[7]{The measurement of $\alpha_{\rm d}$ is done in the following way: The dT laser is switched off $2.8~$ms after the CDT laser, and we let the cloud expand during time-of-flight $t_{\rm ToF}=0.2~$ms before taking an absorption image. At this time, atoms released from the CDT have expanded in the $x-y$ plane more than those released from the dT, which is appropriate for counting each component.}. From figure \ref{evapDimpleShorterNoSimul}(c), we observe that almost all atoms accumulate rapidly (within a few $100~$ms) in the dT. At this stage, the atoms are essentially trapped by the dT in the $x-y$ plane and by the weaker CDT in the $z-$direction. We therefore call this stage ``evaporative filling''. At the end of it, we obtain a cold sodium gas with high phase-space density and collision rate ($\gamma_{\rm coll}\approx2\times10^3$~s$^{-1}$), well suited to start a second evaporative cooling stage.

The difference in trapping frequencies between the cases with and without dimple trap leads to an increase of about $100$ in phase-space density at $t=1~$s. We quantify the evaporation efficiency $\kappa_{\rm evap}$ leading from the starting point ($N_0$,$\mathcal{D}_{0}$) to ($N_1$,$\mathcal{D}_{1}$), using the the definition given in \cite{KetterleDruten_Evaporativecoolingoftrappedatoms},
\begin{equation}
\kappa_{\rm evap} = -\frac{\ln{(\mathcal{D}_1/\mathcal{D}_0})}{\ln{(N_1/N_0)}}.
\end{equation}
Typical values in magnetic traps are $\kappa_{\rm evap}\sim1-2$. In our experiment, we get much better evaporation efficiencies using the dimple trap ($\kappa_{\rm evap}^{dT}\simeq3.5$) than evaporation in the CDT alone ($\kappa_{\rm evap}^{\rm CDT}\simeq1.6$) \footnote[8]{We have found experimentally that turning on the dimple trap at a later stage during the evaporation ramp still results in a boost in phase-space density. However the cooling is not as efficient, so that the final phase-space density and the evaporation efficiency are both slightly worse.}.

We pursue the evaporation by reducing the dT depth, with an exponential ramp from $P_{\rm dT}=200~$mW to $P_{\rm dT}=2~$mW in $1.5~$sec, with a time constant $\tau_{\rm dT}=0.6~$s. This results in a phase-space density increase and a crossing of the BEC threshold after $\sim1~$sec ramping, with $\simeq2\times10^4~$atoms at $T\simeq1~\mu$K. At the end of this ramp, we obtain an almost pure BEC with $\simeq10^4$ atoms.

Finally, we note that the dimple trap is used here in a quite different way compared to the experiment reported in \cite{StamperKurn_ReversibleBEC,Davis_GrowthDynamicBECdimple}. In these works, the authors studied an adiabatic process, in which the gain in phase-space density is obtained isentropically by modifying the trap potential shape \cite{Pinkse_AdiabaticChangingPSD}. In the present work, the entropy is reduced by evaporative cooling as the transfer between the CDT and the dT proceeds.

\section{Conclusion and prospects}

We have demonstrated a method to reach Bose--Einstein condensation of $^{23}$Na in an all-optical experimental setup. We have shown the importance of adapting the trapping potential to the magneto-optical trap cooling dynamics for optimizing the capture in the arms of the crossed trap. In the free evaporation step that follows, an increase in trap depth leads to a fast transfer of the atoms from the arms to the central region, providing thus a dense sample. We have also described the implementation of a two-step evaporation stage using a tightly focused ``dimple'' trap. ``Evaporative filling'' of the dimple trap occurs at almost the same atom number and temperature as in the crossed dipole trap alone. As a result, the phase-space density increases as $\left(\overline{\omega}_{\rm dT}/\overline{\omega}_{\rm CDT}\right)^3$, where $\overline{\omega}_{\rm dT}$ is the dT average frequency and $\overline{\omega}_{\rm CDT}$ the CDT average frequency at low power. Experimentally this corresponds to a large gain in phase-space density, of $\sim100$. After a final evaporation stage in the dimple trap, we are able to obtain almost pure BECs containing $\sim10^4$ atoms.

The efficient ``evaporative filling'' of the dimple trap suggests to generalize the scheme by adding a second, even smaller dimple trap to shorten the time to reach Bose--Einstein condensation. Such a scheme with imbricated evaporative cooling steps (like the layers of an ``atomic matryoshka'') can be taken into consideration if the aim is the production of Bose--Einstein condensates with small atom number, confined in a microscopic potential \cite{Kolomeisky_SingleAtomPipette,Chuu_2005_SubPoissonianNumberStatistics,Dudarev_2007fk_QuantumManyBodyCulling}.

\ack
We wish to thank Lingxuan Shao and Wilbert Kruithof for experimental assistance. D.J. acknowledges financial support by DGA, Contract No. 2008/450. L.D.S. acknowledges ﬁnancial support from the EU IEF grant No. 236240. This work was supported by IFRAF, by the European Union (MIDAS STREP project), and DARPA (OLE project). Laboratoire Kastler Brossel is a Unité Mixte de Recherche (UMR n$^\circ$ 8552) of CNRS, ENS, and UPMC.

\appendix
\section{Analysis of CDT images}
\label{DataAnalysis}

A figure of merit for the loading in the CDT is the atom number in the crossing region of the trap as mainly these atoms will participate to evaporative cooling. We will take the ratio between the atoms in this region and the arms as an indicator of the loading efficiency. We fit the atomic density profiles with a sum of three gaussians, two of them fitting the arms region and the last one fitting the center region,
\begin{equation}
f_{\rm 2D}(x,y)=\sum_{j=1}^{3}G(A_j;x_j,y_j;\sigma_{jx},\sigma_{jy}),
\label{FittingFunction}
\end{equation}
with $G(A,x,y,\sigma_{x},\sigma_{y})=A\rme^{-\frac{1}{2}(x/\sigma_{x})^{2}-\frac{1}{2}(y/\sigma_{y})^{2}}$. Here $A$ is the amplitude, $\sigma_{x}$ and $\sigma_{y}$ the sizes of the distribution along the directions $x$ and $y$. The first two components $1,2$ model the arms so that $\sigma_{1y}\gg \sigma_{1x}$ and $\sigma_{2y}\gg \sigma_{2x}$. The third component models the denser crossing region. The second arm propagates with an angle $\theta$: $(x_2,y_2) = (x_1 \cos(\theta)+y_1 \sin(\theta),-x_1 \sin(\theta)+y_1 \cos(\theta))$. Each arm is supposed to be radially symmetric, the size in the $z$ direction is therefore taken equal to the radial size in the $(x,y)$ plane. 
From the sizes and the calibration of the total fluorescence counts on the CCD with the atom number measured from an absorption image, we infer the atom number in each component $N_1$, $N_2$ and $N_3$. 

In order to evaluate how well equation (\ref{FittingFunction}) can fit the density profile, we perform the fit on a computed density profile of an atomic cloud at thermal equilibrium in a crossed dipole trap potential $U({\bi r})=V_{\rm CDT}({\bi r})$ (see equation (\ref{CDTpotential})), for $P_{\rm CDT}=13.7~$W and $w_{\rm CDT}=42~\mu$m. For a classical gas, the phase-space density $f({\bi r},{\bi p})$ is given by

\begin{equation}
f({\bi r},{\bi p})=\frac{1}{Z}\rme^{-\frac{{\bi p}^2/2M+U({\bi r})}{k_{\rm B} T}}\Theta(-{\bi p}^2/2M-U({\bi r}))
\end{equation}
with $Z$ the partition function chosen such that $\int{f({\bi r},{\bi p})\frac{d^3{\bi r}d^3{\bi p}}{(2\pi\hbar)^3}}=1$, and $\Theta$ the Heavyside step function. An integration of $f({\bi r},{\bi p})$ along the imaging direction $z$ yields a 2D-profile $n_{\rm 2D}^{\rm sim}(x,y)$
\begin{equation}
\label{density}
\fl n_{\rm 2D}^{\rm sim}(x,y)=\int{dz\int{f({\bi r},{\bi p})\frac{d^3{\bi p}}{(2\pi\hbar)^3}}}=\int{dz~\rme^{-U({\bi r})/k_{\rm B} T} \Gamma_{\rm inc}\left(\frac{U({\bi r})}{k_{\rm B} T},\frac{3}{2}\right)},
\end{equation}
where $\Gamma_{\rm inc}$ is the incomplete gamma function. We also calculate the density of states $\rho(\epsilon)$ (with $-V^0_{\rm CDT}<\epsilon<0$)
\begin{equation}
\fl \rho(\epsilon)=\int{\frac{d^3{\bi r}d^3{\bi p}}{(2\pi\hbar)^3}\delta\left(\epsilon-{\bi p}^2/2M-U({\bi r})\right)}=\frac{1}{(2\pi)^2}\left(\frac{2m}{\hbar^2}\right)^{3/2} \int d{\bi r}\sqrt{\epsilon-U({\bi r})}.
\label{dos}
\end{equation}
This can be used to determine the number of atoms $N_{\rm C}$ that have an energy between $-V^0_{\rm CDT}$ and $-V^0_{\rm CDT}/2$,
\begin{equation}
\fl N_{\rm C}=N\left(-V^0_{\rm CDT}\leqslant \epsilon \leqslant  -V^0_{\rm CDT}/2\right)
=n_0\Lambda_{\rm dB}^3\int_{0}^{V^0_{\rm CDT}/2}{d\epsilon~\rho(\epsilon)\rme^{-\epsilon/k_{\rm B} T}},\label{Nc}
\end{equation}
with $n_0$ the density in the center of the trap, and $\Lambda_{\rm dB}=h/\sqrt{2\pi mk_{\rm B}T}$ the thermal de Broglie wavelength. We take $N_{\rm C}$ as an estimate of the number of atoms in the central region. Equations (\ref{dos}) and (\ref{Nc}) are evaluated numerically using Monte-Carlo integration.

\begin{figure}
\includegraphics[scale=1]{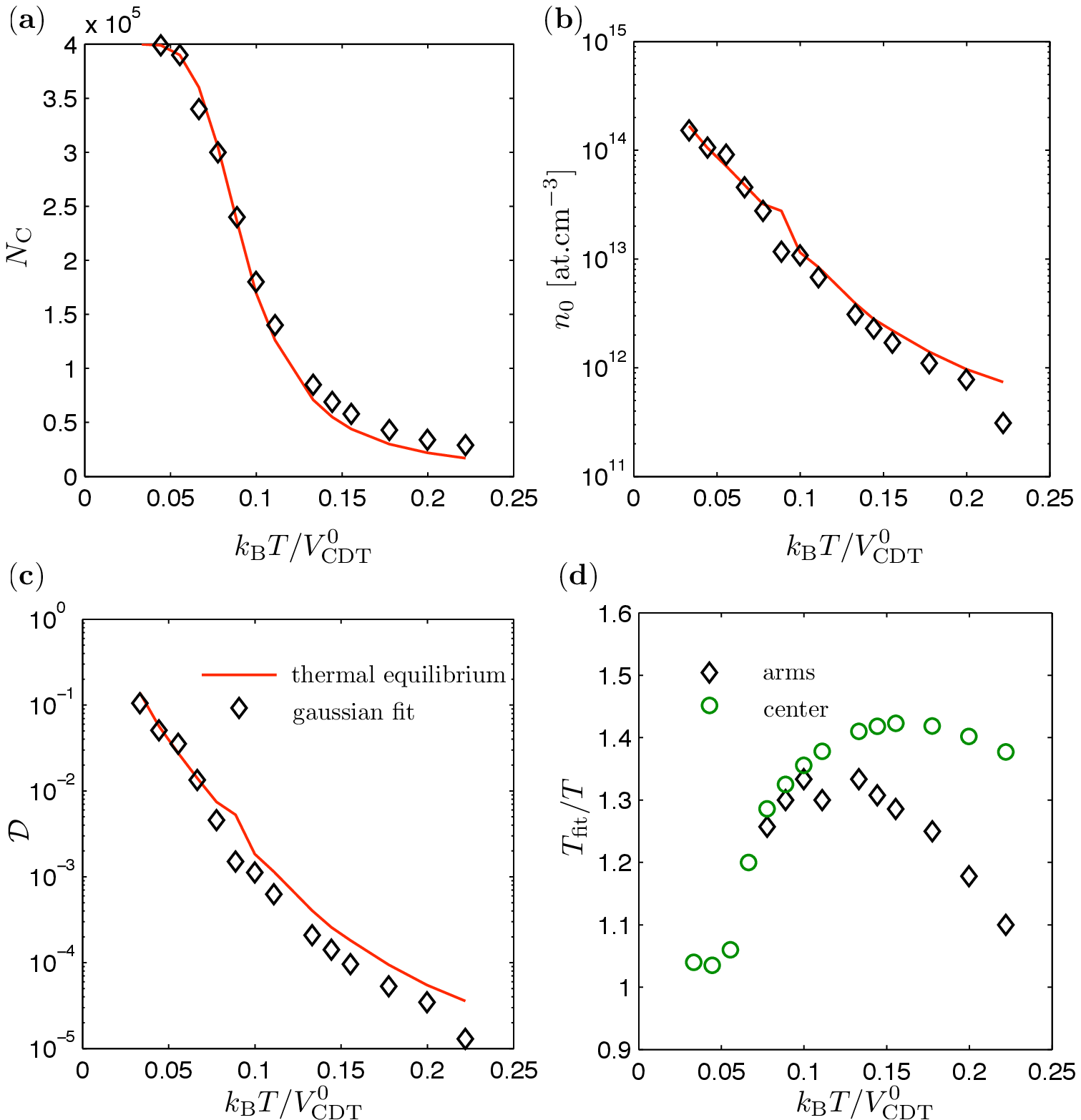}
\caption{Test of the multicomponent fitting routine used to analyze CDT images. For (a), (b) and (c), black diamonds show the result from the fit. In (a), the center atom number is calculated from the density of states and shown as solid line. For the solid lines in (b) and (c), $n_0$ and $\mathcal{D}$ are taken from the formulas at thermal equilibrium. In (d), we compare the temperatures extracted from the size of the arms and the size of the central component to the one used for compute the distribution.}
\label{ComparaisonSimulFit}
\end{figure}

We apply to the simulated profiles $n_{\rm 2D}^{\rm sim}$ the same fitting routine as used for the experimental images. In figure \ref{ComparaisonSimulFit}, we compare the fit output with the parameters used for the simulation.
As one can see, the number of atoms in the center is found to be very close to $N_{\rm C}$. This validates our method to estimate of the loading ratio $\alpha=N_{\rm C}/N_{\rm tot}$ with the result from the fit $N_3/(N_1+N_2+N_3)$. Note however that the procedure systematically over-estimates the temperature in the arms by $\sim30~\%$. This is due to the gaussian shape of the trap that causes a radial density profile wider than the profile created by a harmonic trap with the same curvature. We checked that for a truncated harmonic trap, the fitted temperature is equal to the temperature $T$ obtained for the simulated profile (equation (\ref{density})).\\

\section*{References}
\bibliographystyle{unsrt}
\bibliography{myrefs.bib}	
\end{document}